\documentclass[smallextended,11pt,psfig,a4]{amsart}

\usepackage{amsmath}
\usepackage{amsfonts}
\usepackage{amssymb}
\usepackage{amsthm}
\usepackage{bbm}
\usepackage{MnSymbol}	
\usepackage[dvipsnames]{xcolor}
\usepackage{mathrsfs}

\usepackage[foot]{amsaddr}

\usepackage{geometry}
\usepackage[pdftex]{hyperref}
\usepackage{amssymb}
\usepackage[normalem]{ulem}
\usepackage{xcolor,soul}

\newtheorem{theorem}{Theorem}[section]
\newtheorem{lemma}[theorem]{Lemma}

\newtheorem*{conj*}{Conjecture}

\newtheorem{remark}[theorem]{Remark}

\theoremstyle{definition}

\newtheorem{example}[theorem]{Example}

\theoremstyle{remark}

\numberwithin{equation}{section}

\newcommand{\opn}[1]{\operatorname{#1}}

\newcommand{\mbb}[1]{\mathbb{#1}}

\newcommand{\bs}[1]{\boldsymbol{#1}}

\def\eps{\varepsilon}
\def\>{\rangle}
\def\<{\langle}

\def\Tr{\opn{Tr}}

\def\spn{\opn{span}}

\def\0{\bs{0}}
\def\1{\mathbbm{1}}

\def\C{\mbb{C}}

\def\HH{\mathscr{H}}

\def\Pr{\mathcal{P}}

  \def\XXint#1#2#3{{\setbox0=\hbox{$#1{#2#3}{\int}$}
      \vcenter{\hbox{$#2#3$}}\kern-.47\wd0}}

\begin{document}
\pagestyle{plain}

\def\beq{\begin{equation}}
\def\eeq{\end{equation}}
\def\eps{\epsilon}
\def\laa{\langle}
\def\raa{\rangle}
\def\qed{\begin{flushright} $\square$ \end{flushright}}
\def\qee{\begin{flushright} $\Diamond$ \end{flushright}}
\def\ov{\overline}
\def\bma{\begin{bmatrix}}
\def\ema{\end{bmatrix}}

\def\ora{\overrightarrow}

\def\bma{\begin{bmatrix}}
\def\ema{\end{bmatrix}}
\def\bex{\begin{example}}
\def\eex{\end{example}}
\def\beq{\begin{equation}}
\def\eeq{\end{equation}}
\def\eps{\epsilon}
\def\laa{\langle}
\def\raa{\rangle}
\def\qed{\begin{flushright} $\square$ \end{flushright}}
\def\qee{\begin{flushright} $\Diamond$ \end{flushright}}
\def\ov{\overline}

\author[cfelipe]{C. F. Lardizabal}
\address{Instituto de Matem\'atica e Estat\'istica, Universidade Federal do Rio Grande do Sul, Porto Alegre, RS  91509-900 Brazil.}
\email{cfelipe@mat.ufrgs.br}

\author[lv]{L. Vel\'azquez}
\address{Departamento de Matem\'atica Aplicada \& IUMA, Universidad de Zaragoza, Mar\'ia de Luna 3,
50018 Zaragoza, Spain.}
\email{velazque@unizar.es}

\date{\today}

\title{Mean hitting time formula for positive maps}

\maketitle

\begin{abstract} 
In the classical theory of Markov chains, one may study the mean time to reach some chosen state, and it is well-known that in the irreducible, finite case, such quantity can be calculated in terms of the fundamental matrix of the walk, as stated by the mean hitting time formula. In this work, we present an analogous construction for the setting of irreducible, positive, trace preserving maps. The reasoning on positive maps generalizes recent results given for quantum Markov chains, a class of completely positive maps acting on graphs, presented by S.~Gudder. The tools employed in this work are based on a proper choice of block matrices of operators, inspired in part by recent work on Schur functions for closed operators on Banach spaces, due to F.~A.~Gr\"unbaum and one of the authors. The problem at hand is motivated by questions on quantum information theory, most particularly the study of quantum walks, and provides a basic context on which statistical aspects of quantum evolutions on finite graphs can be expressed in terms of the fundamental matrix, which turns out to be an useful generalized inverse associated with the dynamics. As a consequence of the wide generality of the mean hitting time formula found in this paper, we have obtained extensions of the classical version, either by assuming only the knowledge of the probabilistic distribution for the initial state, or by enlarging the arrival state to a subset of states.
\end{abstract}


\medskip

\section{Introduction}\label{sec1}

In the classical theory of irreducible, finite Markov chains, the mean time of first visit to vertex $j$ starting at $i$ can be calculated as
\begin{equation} \label{mhtf_classica}
E_iT_j = \frac{Z_{jj}-Z_{ji}}{\pi_j}, \qquad Z=(I-\Pr+\Omega)^{-1},
\end{equation}
where $\Pr$ is the stochastic matrix associated with the chain and $\Omega$ is a stochastic matrix whose columns are given by $\pi$, the unique stationary distribution of the walk. Equation (\ref{mhtf_classica}) is called the {\bf mean hitting time formula (MHTF)} \cite{aldous,bremaud} and $Z$, whose columns also add up to one, is called the {\bf fundamental matrix}. A simple proof of such result relies on elementary matrix algebra, a conditioning on the first step reasoning, and establishes a simple method for calculating hitting times in terms of $Z$, which is a generalized inverse of $I-\Pr$. The importance of \eqref{mhtf_classica} relies on its qualitative meaning: the mean time to reach any state $j$ starting from a state $i$ is determined by the case $i=j$ --from which one has the mean return time for the state $j$, given by $1/\pi_j$ according to Kac's lemma  \cite{durrett,kac}-- combined with the fundamental matrix.  

\medskip

With the recent interest in the study of quantum versions of walks on graphs, a natural question in the context of hitting probabilities and mean times of first visit is to ask whether a similar formula is available in a quantum setting. 
Recently, a version of the mean hitting time formula has been presented in the setting of Open Quantum Walks (OQWs) \cite{attal,oqwmhtf1} and this was later generalized for Quantum Markov Chains (QMCs) \cite{oqwmhtf2}. QMCs consist of an important class of completely positive maps described by S.~Gudder \cite{gudder} and their statistical behavior is typically associated with dissipative (open) quantum dynamics on graphs. A natural question is: can such a result be extended to more general positive maps? 

\medskip

In this work, we present a version of the mean hitting time formula for positive, irreducible, trace preserving maps on finite dimensional Hilbert spaces.

\medskip

The result for positive maps is stated by theorems \ref{mhtf} and \ref{mhtfg}. The proof is self-cointained and extends the results given by \cite{oqwmhtf1,oqwmhtf2} in two directions. First, one is now able to consider {\it any} irreducible, positive, trace preserving map, as the proof generalizes the reasoning presented in \cite{oqwmhtf1}, which relies strongly on the block matrix description of the OQW. As such block structure is usually absent for general positive maps, one has to rely solely on the projection operators associated with the initial and final states of interest. Second, the theorems concern the hitting time to subspaces of {\it any} finite dimension, generalizing the statement seen in \cite{oqwmhtf2}, which only considered the case of mean hitting time to a site (i.e., a vertex together with its internal degree of freedom).

\medskip

Just as in the classical case, we do not claim that to obtain mean hitting times in terms of the fundamental matrix is easier than via direct calculation, i.e., by definition: both problems involve the practical problem of calculating matrix inverses. The result shows that, similarly to the classical case, the fundamental matrix combined with information about return times encodes all the information about mean hitting times. 

\medskip

The problem of obtaining versions of the formulae presented in this work for the context of infinite-dimensional Hilbert spaces, together with an associated spectral analysis, will be the topic of a future work. We note that such a project is in part motivated by the recurrence theory of unitary steps developed recently in terms of Schur functions \cite{bourg, gvfr, werner}.

\medskip

The contents of this work are organized as follows. Section~\ref{sec2} introduces a generalization of the fundamental matrix for irreducible, positive, trace preserving maps and studies its properties. In Section~\ref{sec3} some probabilistic notions are defined, relating them with certain block matrices needed later. The MHTF for positive maps is obtained in Section~\ref{sec4}, including illustrative examples. Finally, Section~\ref{sec5} connects the above MHTF with the classical one \eqref{mhtf_classica}, uncovering also some classical generalizations provided by the results of this work.

\section{The fundamental map}\label{sec2}

Let $\mathbb{C}$ denote the set of complex numbers and $M_n=M_n(\mathbb{C})$ the set of order $n$ complex matrices. Throughout this work we will assume $n<\infty$. Denoting by $X^*$ the complex conjugate of the transpose $X^T$ of a matrix $X$, we say that $X\in M_n$ is positive semidefinite (or just positive), denoted by $X\geq 0$, whenever $v^*Xv\ge0$ for every $v\in\mathbb{C}^n$. If $v^*Xv>0$ for every $v\in\mathbb{C}^n\setminus\{0\}$ we say that $X$ is strictly positive, written as $X>0$. A linear map $\Phi:M_n\to M_n$ is called {\bf positive} whenever $X\geq 0$ implies $\Phi(X)\geq 0$. A positive map $\Phi$ is {\bf trace preserving} if $\Tr(\Phi(X))=\Tr(X)$. A {\bf state or density matrix} is a positive matrix $X$ such that $\Tr(X)=1$.

\medskip

We say that the positive map $\Phi:M_n\to M_n$ is {\bf irreducible} if the only orthogonal projections $P$ on $M_n$ such that $\Phi$ leaves $P M_nP$ globally invariant --that is, $\Phi(P M_nP)\subset  PM_nP$-- are $P=0$ and $P=I_n$, where $I_n$ is the order $n$ identity. Every positive trace preserving map $\Phi$ on $M_n$ has an {\bf invariant state} $\pi$, i.e. a density matrix such that $\Phi(\pi)=\pi$. If, besides, $\Phi$ is irreducible, then $\pi>0$ and any other fixed point $X\in M_n$ of $\Phi$ is a multiple of $\pi$, i.e. $\Phi(X)=X$ implies $X=\lambda\pi$, $\lambda\in\C$ \cite{evans}.

\medskip

If $A\in M_n$, the corresponding vector representation $vec(A)$ associated with it is given by stacking together its matrix rows. For instance, if $n=2$,
$$
A = \begin{bmatrix} A_{11} & A_{12} \\ A_{21} & A_{22}\end{bmatrix}
\quad \Rightarrow \quad 
vec(A) = \begin{bmatrix} A_{11} & A_{12} & A_{21} & A_{22}\end{bmatrix}^T.
$$
The $vec$ mapping satisfies $vec(AXB^T)=(A\otimes B)\,vec(X)$ for any square matrices $A, B, X$ \cite{hj2}. In particular, $vec(BXB^*)=vec(BX\ov{B}^T)=(B\otimes \ov{B})\,vec(X)$. This allows us to consider the {\bf matrix representation} of any map $\Phi$ acting on $M_n$: we define $\lceil\Phi\rceil$ to be the matrix such that $\Phi(X)=vec^{-1}(\lceil\Phi\rceil \, vec(X))$ for every $X\in M_n$. This matrix representation is very useful when performing calculations in concrete examples (see Examples~\ref{ex1} and \ref{ex2}). 

\medskip

The $vec$ mapping establishes a unitary equivalence between the Hilbert spaces $M_n$ and $\C^{n^2}$ with the corresponding standard inner products, $\langle \, \cdot \mid \cdot \, \rangle_{M_n}$ and $\langle \, \cdot \mid \cdot \, \rangle_{\C^{n^2}}$, because
$$
\langle B | A \rangle_{M_n} = \Tr(B^*A) = \sum_{i,j}\overline{B_{ij}}A_{ij} = vec(B)^*vec(A)
= \langle vec(B) | vec(A) \rangle_{\C^{n^2}},
\qquad A,B\in M_n.
$$
When no misunderstanding may arise we will use the same notation $\langle \, \cdot \mid \cdot \, \rangle$ for the inner products in $M_n$ and $\C^n$.

\medskip

The adjoint of a linear map $\Phi \colon M_n \to M_n$ with respect to the above inner product in $M_n$ is the linear map $\Phi^* \colon M_n \to M_n$ defined by
$$
\langle \Phi^*(Y)|X \rangle = \langle Y|\Phi(X) \rangle,
\qquad X,Y\in M_n.
$$
It is obvious that their matrix representations are related by $\lceil\Phi^*\rceil=\lceil\Phi\rceil^*$. In terms of its adjoint, the trace preserving condition for $\Phi$ reads as $\Phi^*(I_n)=I_n$. 

\medskip

Given $X,Y\in M_n$, the inner product in $M_n$ allows us to define the ket-bra linear map $|X \rangle \langle Y| \colon M_n \to M_n$, 
$$
|X \rangle \langle Y|A = \langle Y|A \rangle X,
\qquad A\in M_n,
$$
so that $\Phi|X\rangle\langle Y| = |\Phi(X)\rangle\langle Y|$ and $|X\rangle\langle Y|\Phi = |X\rangle\langle\Phi^*(Y)|$. 

\medskip
 
Any matrix $A\in M_n$ may be viewed as an operator $\phi \mapsto A\phi$ on the Hilbert space $\HH_n=\C^n$ with the standard inner product. In this space, ket-bra operators also make sense: any two vectors $\phi,\psi\in\HH_n$ generate the ket-bra operator $|\phi\rangle\langle\psi|=\phi\psi^*\in M_n$. The unit vectors $\phi\in\HH_n$ are called {\bf pure states} since they may be identified up to an arbitrary phase with the states $\rho_\phi:=|\phi\rangle\langle\phi|=\phi\phi^*$. We will refer to $\HH_n$ as the {\bf pure state Hilbert space} for $M_n$.

\medskip

The first step to obtain a MHTF for positive maps is to introduce the analogue of the fundamental matrix for such maps. For this purpose, one may look at the fundamental matrix of an irreducible Markov chain with stochastic matrix $\Pr$ in the following way,
$$
 Z = (I - \Pr + \Omega)^{-1}, \qquad \Omega = \pi \ \mathbf{1}^* = |\pi\rangle\langle\mathbf{1}|,
 \qquad \mathbf{1} = \begin{pmatrix} 1 \\ 1 \\ \vdots \\ 1 \end{pmatrix},
$$
where $\pi$ is the state spanning the subspace of right eigenvectors of $\Pr$ with eigenvalue 1, while $\mathbf{1}^*$ turns out to be a left eigenvector of $\Pr$ with eigenvalue 1 since this matrix is stochastic. In other words, $\pi$ and $\mathbf{1}$ are fixed points for the transformations defined by the matrices $\Pr$ and $\Pr^*$ respectively. 

\medskip

Therefore, a natural generalization of this fundamental matrix for a positive, irreducible, trace preserving map $\Phi$ on $M_n$ should be given by the map
$$
 Z = (I - \Phi + \Omega)^{-1}, \qquad \Omega = |\pi\rangle\langle I_n|,
$$
where $I$ is the identity map on $M_n$, $\pi$ is the unique invariant state of $\Phi$ and the order $n$ identity matrix $I_n$ is a fixed point for $\Phi^*$, as follows from the trace preserving condition for $\Phi$. Here, the ket-bra operator $\Omega$ is understood with respect to the standard inner product in $M_n$. Since this inner product makes $M_n$ unitarily equivalent to the canonical Hilbert space $\C^{n^2}$ via $vec$ transformation, using $vec$ notation the fundamental matrix reads as
$$
 \lceil Z \rceil = (I_{n^2} - \lceil \Phi \rceil + \lceil \Omega \rceil)^{-1}, 
 \qquad 
 \lceil \Omega \rceil = vec(\pi) \ vec(I_n)^T.
$$ 

The existence of this extension of the fundamental matrix is guaranteed by the following lemma.

\begin{lemma} (Existence of the fundamental map). \label{FM}

Let $\Phi$ be an irreducible, trace preserving, positive map on $M_n$ with invariant state $\pi$. Then $I-\Phi+\Omega$ is invertible, where $\Omega=|\pi\rangle\langle I_n|$ is a trace preserving map satisfying
\begin{equation} \label{propOmega}
 \Omega^2 = \Phi\Omega = \Omega\Phi = \Omega.
\end{equation} 
We will refer to $Z=(I-\Phi+\Omega)^{-1}$ as the {\bf fundamental map} of $\Phi$.
\end{lemma}

{\bf Proof.} 
Note that $\Omega$ is trace preserving since $\Omega(\rho) = \langle I|\rho\rangle\,\pi = \Tr(\rho)\,\pi$ and $\Tr(\pi)=1$ because $\pi$ is a state. Besides, $\Omega^2 = |\pi\rangle\langle I|\pi\rangle\langle I| = \Tr(\pi)\,\Omega = \Omega$. Also, $\Phi\Omega = |\Phi(\pi)\rangle\langle I| = \Omega$ since $\pi$ is invariant for $\Phi$, while $\Omega\Phi = |\pi\rangle\langle \Phi^*(I)| = \Omega$ because $\Phi$ is trace preserving. Regarding the invertibility of $I-\Phi+\Omega$, it is enough to prove that, assuming that $X\in M_n$ satisfies $(I-\Phi+\Omega)(X)=0$, then $X=0$. The above properties show that $\Omega=\Omega(I-\Phi+\Omega)$, which leads to $\Omega(X) = \Omega(I-\Phi+\Omega)(X)=0$. This implies that $(I-\Phi)(X)=0$, hence $X=\lambda\pi$, $\lambda\in\mathbb{C}$, due to the essential uniqueness of the fixed point for $\Phi$. Thus, $\Omega(X)=\lambda\Tr(\pi)\,\pi=\lambda\pi$, concluding that $\lambda=0$ and $X=0$.   
\qed

The properties of the operator $\Omega$ given in the previous lemma are the origin of useful properties of the fundamental map.

\begin{lemma} \label{omega1lema} (Properties of the fundamental map). 
The fundamental map $Z=(I-\Phi+\Omega)^{-1}$ of an irreducible, trace preserving, positive map $\Phi$ is also trace preserving and satisfies
\beq\label{mass1}  {Z} {\Omega}= {\Omega} {Z}= {\Omega},\eeq
\beq\label{mass2}  {Z}(I- {\Phi})=I- {Z} {\Omega}=I- {\Omega},\eeq
\beq\label{mass3}  (I- {\Phi}) {Z}=I- {\Omega} {Z}=I- {\Omega}.\eeq
\end{lemma}

{\bf Proof.} 
$\Phi$ and $I-\Phi+\Omega$ are simultaneously trace preserving because $I$ and $\Omega$ are always trace preserving. This proves that $Z=(I-\Phi+\Omega)^{-1}$ is trace preserving whenever it exists. From \eqref{propOmega} we find that
$$
\begin{aligned} 
& {Z} (I- {\Phi}+ {\Omega}) \Omega = \Omega 
\quad\Rightarrow\quad  
{Z}{\Omega}- {Z}\underbrace{{\Phi}{\Omega}}_{{\Omega}}+ {Z}\underbrace{{\Omega}^2}_{{\Omega}} 
= {\Omega}
\quad\Rightarrow\quad 
{Z} {\Omega}= {\Omega},
\\
& \Omega (I- {\Phi}+ {\Omega}) {Z} = \Omega
\quad\Rightarrow\quad  
{\Omega}{Z}- \underbrace{{\Omega}{\Phi}}_{{\Omega}}{Z}+ \underbrace{{\Omega}^2}_{{\Omega}}{Z} 
= {\Omega}
\quad\Rightarrow\quad
{\Omega} {Z}= {\Omega}.
\end{aligned}
$$
Also,
$$
\begin{aligned}  
{Z}(I- {\Phi}+ {\Omega})=I
\quad\Rightarrow\quad  
{Z}(I- {\Phi})=I- {Z} {\Omega}=I- {\Omega},
\\
(I- {\Phi}+ {\Omega}) {Z}=I
\quad\Rightarrow\quad
(I- {\Phi}) {Z}=I- {\Omega} {Z}=I- {\Omega}.
\end{aligned}
$$
\qed

\section{Probabilistic notions for positive maps} \label{sec3}

Born's rule gives to measurements a prominent role in quantum mechanics, absent in the classical case. In consequence, introducing in discrete time quantum mechanics certain probabilistic notions such as hitting times requires the choice of the experimental setting specifying the measurements needed to define such notions. We adopt here the so called {\bf monitoring approach} in which a measurement is performed after each evolution step to check whether the system has reached a state lying in a prescribed target subspace $V$ or not (see \cite{bach,bourg,gvfr,werner,brun,gawron,sinkovicz}). For this purpose, we need to associate to $V$ certain orthogonal projections conditioning on the events ``the system is in a state lying in $V$" or its complementary. 

\medskip

Let $V$ denote a subspace of the pure state Hilbert space $\HH_n$ where the operators represented by the matrices of $M_n$ act. If $P$ is the orthogonal projection onto $V$ and $Q=I_n-P$, we introduce the corresponding orthogonal projections, $\mathbb{P}$ and $\mathbb{Q}$, for the space $M_n$ where the mixed states live,
$$
\mathbb{P},\mathbb{Q}:M_n\to M_n, \qquad \mathbb{P}=P\cdot P, \qquad \mathbb{Q}=Q\cdot Q.
$$
Note  that $\mathbb{P}+\mathbb{Q}+\mathbb{R}=I$, where $\mathbb{R}:=P\cdot Q+Q\cdot P$ is an orthogonal projection onto traceless matrices.

\medskip

Given a pure state $\phi\in\HH_n$, we define the following probabilistic quantities for a discrete time process whose steps follow by iterating a positive trace preserving map $\Phi$ on $M_n$:
\begin{equation} \label{prob}
\begin{aligned}
& p_r(\phi \to V) = & & \text{probability of reaching the subspace $V$ in $r$ steps when starting at $\phi$.} 
\\
& \pi_r(\phi\to V) = & & \text{probability of reaching the subspace $V$ for the first time in $r$ steps} 
\\
& & & \text{when starting at $\phi$.} 
\\
& \pi(\phi\to V) = & & \text{probability of ever reaching the subspace $V$} 
& & \kern-100pt \text{\bf(hitting probability)} 
\\
& & & \text{when starting at $\phi$.}  
\\
& \tau(\phi\to V) = & & \text{expected time of first visit to the subspace $V$}
& & \kern-100pt \text{\bf(mean hitting time)}
\\
& & & \text{when starting at $\phi$.}  
\end{aligned}
\end{equation}
We will refer to $\phi$ and $V$ as the {\bf initial state} and the {\bf arrival subspace} respectively. An important particular case of the subspace $V$ is the 1-dimensional space generated by a pure state $\psi\in\HH_n$. In this case we will simplify the notation by writing $\pi(\phi\to \psi)$, $\tau(\phi\to \psi)$, and so on.   

\medskip

Another particularly interesting situation arises when $\phi\in V$. This defines the so called {\bf recurrence properties} of the process governed by the map $\Phi$, for instance, the probability $\pi(\phi\to V)$ of the return of $\phi$ to the subspace $V$ {\bf (return probability)}, and the expected time $\tau(\phi\to V)$ of such a return {\bf (mean return time)}.

\medskip

According to the monitoring approach, the above notions can be expressed in terms of appropriate trace expressions involving maps on $M_n$ acting on the density matrix $\rho_\phi=|\phi\>\<\phi|$ of the initial state:
\begin{flalign}
& p_r(\phi\to V) = \Tr(\mathbb{P}\Phi^r \rho_\phi), 
\nonumber \\
& \pi_r(\phi\to V) = \Tr(\mathbb{P}\Phi(\mathbb{Q}\Phi)^{r-1} \rho_\phi), 
\nonumber \\
& \pi(\phi\to V) = \sum_{r\geq 1} \pi_r(\phi\to V) = 
\sum_{r\geq 1} \Tr(\mathbb{P}\Phi(\mathbb{Q}\Phi)^{r-1} \rho_\phi), 
\label{probPOS}  \\
&\tau(\phi\to V)=
\begin{cases}
\infty, 
& \text{ if } \pi(\phi\to V)<1,  
\\
\displaystyle \sum_{r\geq 1} r\pi_r(\phi\to V) = 
\sum_{r\geq 1} r\Tr(\mathbb{P}\Phi(\mathbb{Q}\Phi)^{r-1} \rho_\phi), 
& \text{ if } \pi(\phi\to V)=1. 
\end{cases} 
\nonumber 
\end{flalign}
Here and in what follows we write for convenience $\Phi\rho$ to denote $\Phi(\rho)$ for any $\rho\in M_n$. Since $I-\mathbb{Q}=\mathbb{P}+\mathbb{R}$ differs from $\mathbb{P}$ in an orthogonal projection $\mathbb{R}$ onto traceless matrices, the traces appearing in \eqref{probPOS} may be equivalently expressed by substituting $\mathbb{P}$ by $I-\mathbb{Q}$. This will be key for the rest of the paper.

\medskip

Given an irreducible, trace preserving, positive map $\Phi$ on $M_n$, it turns out that $\pi(\phi\to V)=1$ for every pure state $\phi$ and every subspace $V$ (see \cite{glv} and Remark~\ref{finrec} below). Therefore, in this case the expression of the mean hitting time is the one in the bottom line of \eqref{probPOS}.

\medskip

In this context, the main question we intend to answer is about the existence of a MHTF for processes governed by positive irreducible trace preserving maps. This should take the form of a relation between the above introduced mean hitting time and the fundamental map given in Lemma~\ref{FM}.

\medskip
 
In order to address the above problem we introduce the function
$$
\mathbb{G}(z) = 
\sum_{r\geq 1} \Phi(\mathbb{Q}\Phi)^{r-1}z^r = 
z\Phi (I-z\mathbb{Q}\Phi)^{-1}, 
\qquad |z|<1,
$$
which is analytical because, with respect to the operator norm, $\|\mathbb{Q}\|=1$ as an orthogonal projection, while the Russo-Dye theorem  \cite{bhatia} implies that $\|\Phi\|=\|\Phi^*\|=\|\Phi^*(I_n)\|=\|I_n\|=1$ since $\Phi$ is trace preserving. The interest of this function relies on the fact that, according to \eqref{probPOS}, its boundary behaviour around $z=1$ provides hitting probabilities and mean hitting times,
\begin{equation} \label{HKlim}
\begin{aligned} 
& \pi(\phi\to V) = \Tr(\mathbb{P}H\rho_\phi) = \Tr((I-\mathbb{Q})H\rho_\phi),
& \qquad & H:=\lim_{x\uparrow 1}\mathbb{G}(x),
\\
& \tau(\phi\to V) = \Tr(\mathbb{P}K\rho_\phi) = \Tr((I-\mathbb{Q})K\rho_\phi),
& & K:=\lim_{x\uparrow 1}\mathbb{G}'(x).
\end{aligned}
\end{equation}
As limits of positive maps, $H$ and $K$ are positive maps too. We will refer to $H$ and $K$ as the {\bf hitting probability map} and the {\bf mean hitting time map} for the subspace $V$. The existence of these limits is guaranteed by the following remark.

\begin{remark} \label{finrec}
Since the pure state space $\HH_n$ has finite-dimension, it is known that 1 does not lie in the spectrum of $\mathbb{Q}\Phi$ for any irreducible, trace preserving, positive map $\Phi$ and any non-trivial orthogonal projection $Q$ \cite[proof of Theorem 2.8]{glv} (the proof given there for completely positive maps also holds for positive maps because it completely relies on the existence, uniqueness and strictly positiveness of an invariant state guaranteed by the irreducibility and the trace preserving condition). This ensures the analyticity of $\mathbb{G}(z)$ at $z=1$, which allows us to express  
\begin{equation} \label{HK}
\begin{aligned}
& H = \mathbb{G}(1) = \Phi (I-\mathbb{Q}\Phi)^{-1},
\\
& K = \mathbb{G}'(1) = 
\Phi (I-\mathbb{Q}\Phi)^{-1} + 
\Phi \mathbb{Q} \Phi (I-\mathbb{Q}\Phi)^{-2} =
\Phi (I-\mathbb{Q}\Phi)^{-2}.
\end{aligned}
\end{equation}
From the first equality, and bearing in mind that $\Phi$ is trace preserving, we find that the map $(I-\mathbb{Q})H$ giving the hitting probability is also trace preserving,
$$
\Tr((I-\mathbb{Q})H\rho) 
= \Tr(H\rho) - \Tr(\mathbb{Q}H\rho) 
= \Tr((I-\mathbb{Q}\Phi)^{-1}\rho) - \Tr(\mathbb{Q}\Phi(I-\mathbb{Q}\Phi)^{-1}\rho)
= \Tr(\rho),
\qquad
\rho\in M_n.
$$
Hence, for any pure state $\phi$ and any subspace $V$ of pure states,
$$
\pi(\phi\to V) = \Tr((I-\mathbb{Q})H\rho_\phi) = \Tr(\rho_\phi) = 1.
$$
That is, any hitting probability equals 1 for an irreducible, trace preserving, positive map on $M_n$. Besides, in such a case the mean hitting time is finite and may be expressed as
$$
\begin{aligned}
\tau(\phi\to V)
& = \Tr((I-\mathbb{Q})K\rho_\phi) 
= \Tr(K\rho_\phi) - \Tr(\mathbb{Q}K\rho_\phi)
= \Tr((I-\mathbb{Q}\Phi)^{-2}\rho_\phi) - \Tr(\mathbb{Q}\Phi(I-\mathbb{Q}\Phi)^{-2}\rho_\phi)
\\
& = \Tr((I-\mathbb{Q}\Phi)^{-1}\rho_\phi) = \Tr(H\rho_\phi).
\end{aligned} 
$$
\end{remark}

\medskip

The next step is to consider the following $V$-dependent {\bf block matrix representation} of a map $T \colon M_n \to M_n$:
\beq\label{phirep}
T = 
\begin{bmatrix} T_{11} & T_{12} \\ T_{21} & T_{22}\end{bmatrix} :=
\begin{bmatrix} 
(I-\mathbb{Q})T(I-\mathbb{Q}) & (I-\mathbb{Q})T\mathbb{Q} \\ 
\mathbb{Q}T(I-\mathbb{Q}) & \mathbb{Q}T\mathbb{Q}
\end{bmatrix}.
\eeq

\medskip

The block matrix representation of the positive maps $H$ and $K$ yields information about the first visit to the subspace $V$, when starting at states supported on $V$ or $V^\bot$. More precisely, from \eqref{HKlim} we find that
\begin{align} \label{piH}
& \begin{aligned}
& \pi(\phi\to V) = 
\begin{cases} 
\Tr((I-\mathbb{Q})H(I-\mathbb{Q})\rho_\phi) = \Tr(H_{11}\rho_\phi), 
& \text{if} \; \phi\in V, 
\\
\Tr((I-\mathbb{Q})H\mathbb{Q}\rho_\phi) = \Tr(H_{12}\rho_\phi), 
& \text{if} \; \phi\in V^\bot,
\end{cases}
\qquad H=\begin{bmatrix} H_{11} & H_{12} \\ H_{21} & H_{22} \end{bmatrix},
\end{aligned}
\\ \label{tauK}
& \begin{aligned}
& \tau(\phi\to V) = 
\begin{cases} 
\Tr((I-\mathbb{Q})K(I-\mathbb{Q})\rho_\phi) = \Tr(K_{11}\rho_\phi), 
& \text{if} \; \phi\in V, 
\\
\Tr((I-\mathbb{Q})K\mathbb{Q}\rho_\phi) = \Tr(K_{12}\rho_\phi), 
& \text{if} \; \phi\in V^\bot,
\end{cases}
\qquad K=\begin{bmatrix} K_{11} & K_{12} \\ K_{21} & K_{22} \end{bmatrix}.
\end{aligned}
\end{align}
In particular, the first diagonal blocks of $H$ and $K$ provide recurrence properties: $H_{11}$ and $K_{11}$ give respectively the probabilities and mean times for the return of a state $\phi\in V$ to the subspace $V$ where it lies. We will call $H_{11}$ and $K_{11}$ the {\bf return probability map} and the {\bf mean return time map} for the subspace $V$, respectively.

\medskip

The second row of $H$ and $K$ also yield information about the transitions $\phi \to V$. For example, from Remark~\ref{finrec} one finds that $\Tr(\mathbb{Q}H\rho_\phi) = \Tr(H\rho_\phi)-\Tr((I-\mathbb{Q})H\rho_\phi) = \tau(\phi\to V) - 1$. Therefore,
$$
\tau(\phi\to V) - 1 =  
\begin{cases} 
\Tr(\mathbb{Q}H(I-\mathbb{Q})\rho_\phi) = \Tr(H_{21}\rho_\phi), 
& \text{if} \; \phi\in V, 
\\
\Tr(\mathbb{Q}H\mathbb{Q}\rho_\phi) = \Tr(H_{22}\rho_\phi), 
& \text{if} \; \phi\in V^\bot.
\end{cases}
$$

\medskip
  
The idea of employing block matrices of operators in the context of first visit/return to subspaces has been illustrated, for instance, in \cite{gvfr,oqwmhtf1,oqwmhtf2}. Such block matrices play a main role in the deduction of the MHTF for positive maps given in the next Section.

\section{MHTF for positive maps}\label{sec4}

Let $\Phi$ denote an irreducible, positive, trace preserving map acting on $M_n$, and let $\pi$ be the unique strictly positive density matrix which is invariant for $\Phi$ \cite{evans}. This section will establish a MHTF for $\Phi$ involving the fundamental map
\begin{equation} \label{FMPM}
Z=(I-\Phi+\Omega)^{-1}, \qquad \Omega=|\pi\rangle\langle I_n|. 
\end{equation} 
This MHTF will relate such a fundamental map with the mean time $\tau(\phi\to V)$ of the first visit of a pure state $\phi$ to a subspace $V$ of pure states. The subspace $V$ will enter into the MHTF via the corresponding mean return time map $K_{11}$. 

\medskip

We start by introducing the following maps, defined in terms of the block matrix representation of $K$:
\begin{equation} \label{DKL} 
{D}:= \begin{bmatrix} K_{11} & 0 \\ 0 & K_{22} \end{bmatrix},
\quad
{N}:= {K}-{D} = \begin{bmatrix} 0 & K_{12} \\ K_{21} & 0 \end{bmatrix},
\quad 
{L}:= {K}-{N} {\Phi} =
\begin{bmatrix} 
K_{11}-K_{12}\Phi_{21} & K_{12}-K_{12}\Phi_{22} 
\\ 
K_{21}-K_{21}\Phi_{11} & K_{22}-K_{21}\Phi_{12}
\end{bmatrix}.
\end{equation}
The next lemma summarizes some relations among the above maps and the fundamental map $Z$.

\begin{lemma}\label{lema43}
For every $\psi\in V$ and $\phi\in V^\perp$ we have that
$$
N_{12}\rho_\phi=(DZ)_{11}\rho_\psi-(DZ)_{12}\rho_\phi+[(LZ)_{12}\rho_\phi-(LZ)_{11}\rho_\psi],
$$
$$
N_{21}\rho_\psi=(DZ)_{22}\rho_\phi-(DZ)_{21}\rho_\psi+[(LZ)_{21}\rho_\psi-(LZ)_{22}\rho_\phi].
$$
\end{lemma}

{\bf Proof.} 
From the definition of the maps it follows that
$$ 
{N} {\Phi} + {L} = {N}+ {D}
\quad\Rightarrow\quad  
{N} {\Phi} {Z} + {L} {Z} = {N} {Z} + {D} {Z}
\quad\Rightarrow\quad  
{N}(I-{\Phi}) {Z} = {L} {Z} - {D} {Z}.
$$
Combining this with ${N}(I-{\Phi}) {Z} = {N}(I-{\Omega}) = {N} - {N} {\Omega}$, which follows from \eqref{mass3}, yields 
\begin{equation} \label{eqNLDZ} 
{N} - {N} {\Omega} = {L} {Z} - {D} {Z}
\quad\Rightarrow\quad  
{N} = {L} {Z} - {D} {Z} + {N} {\Omega}.
\end{equation}
Now note that, in terms of block matrix representations,
$$
{N}{\Omega} = 
\begin{bmatrix} 0 & K_{12} \\ K_{21} & 0 \end{bmatrix}
\begin{bmatrix} 
(I-\mathbb{Q})\Omega(I-\mathbb{Q}) & (I-\mathbb{Q})\Omega \mathbb{Q} 
\\ 
\mathbb{Q}\Omega(I-\mathbb{Q}) & \mathbb{Q}\Omega\mathbb{Q} 
\end{bmatrix}
= \begin{bmatrix} 
K_{12}\mathbb{Q}\Omega(I-\mathbb{Q}) & K_{12}\mathbb{Q}\Omega\mathbb{Q} 
\\ 
K_{21}(I-\mathbb{Q})\Omega(I-\mathbb{Q}) & K_{21}(I-\mathbb{Q})\Omega\mathbb{Q}
\end{bmatrix},
$$
while the diagonal blocks of \eqref{eqNLDZ} give
\begin{align}
& 0=(LZ)_{11}-(DZ)_{11}+(N\Omega)_{11}
\quad\Rightarrow\quad 
(N\Omega)_{11}\rho_\psi=[(DZ)_{11}-(LZ)_{11}]\rho_\psi,
\label{tri1} \\ 
& 0=(LZ)_{22}-(DZ)_{22}+(N\Omega)_{22}
\quad\Rightarrow\quad 
(N\Omega)_{22}\rho_\phi=[(DZ)_{22}-(LZ)_{22}]\rho_\phi.
\label{tri2}
\end{align}
Since $\Omega\rho = \langle I_n|\rho\rangle\,\pi = \Tr(\rho)\, \pi = \pi$ for any state $\rho$, we conclude that $\Omega\rho_\psi=\Omega\rho_\phi$. Hence,
$$
(N\Omega)_{11}\rho_\psi = K_{12}\mathbb{Q}\Omega (I-\mathbb{Q})\rho_\psi = 
K_{12}\mathbb{Q}\Omega\rho_\psi = K_{12}\mathbb{Q}\Omega\rho_\phi = 
K_{12}\mathbb{Q}\Omega \mathbb{Q}\rho_\phi = (N\Omega)_{12}\rho_\phi,
$$
and analogously for the second row of $N\Omega$. We obtain
\beq\label{tri3}(N\Omega)_{11}\rho_\psi=(N\Omega)_{12}\rho_\phi   ,\;\;\; (N\Omega)_{21}\rho_\psi=(N\Omega)_{22}\rho_\phi.\eeq
Then, by using the first equality in (\ref{tri3}) followed by (\ref{tri1}), the non-diagonal upper right block of \eqref{eqNLDZ} leads to
$$
\begin{aligned} 
N_{12}\rho_\phi & = [(LZ)_{12}-(DZ)_{12}+(N\Omega)_{12}]\rho_\phi 
= [(LZ)_{12}-(DZ)_{12}]\rho_\phi + (N\Omega)_{11}\rho_\psi
\\
& = [(LZ)_{12}-(DZ)_{12}]\rho_\phi+[(DZ)_{11}-(LZ)_{11}]\rho_\psi,
\end{aligned}
$$
that is,
\beq\label{lres1}
N_{12}\rho_\phi = (DZ)_{11}\rho_\psi-(DZ)_{12}\rho_\phi+[(LZ)_{12}\rho_\phi-(LZ)_{11}\rho_\psi].
\eeq
Similarly, by using the second equality in (\ref{tri3}) followed by (\ref{tri2}), the non-diagonal lower left block of \eqref{eqNLDZ} yields
$$
\begin{aligned}
N_{21}\rho_\psi & = [(LZ)_{21}-(DZ)_{21} +(N\Omega)_{21}]\rho_\psi 
= [(LZ)_{21}-(DZ)_{21}]\rho_\phi+(N\Omega)_{22}\rho_\phi
\\
& = [(LZ)_{21}-(DZ)_{21}]\rho_\psi+[(DZ)_{22}-(LZ)_{22}]\rho_\phi,
\end{aligned}
$$
so that,
\beq\label{lres2}
N_{21}\rho_\psi = (DZ)_{22}\rho_\phi-(DZ)_{21}\rho_\psi+[(LZ)_{21}\rho_\psi-(LZ)_{22}\rho_\phi].\eeq

\qed

The previous result will be key to obtain the MHTF for irreducible, positive, trace preserving maps. However, for that purpose, we also need to show that the first row of the block matrix representation of $L$ is trace preserving too.

\begin{lemma} \label{Ltr}
The first block matrix rows of $L$ and $H$ coincide, i.e. $(I-\mathbb{Q})L=(I-\mathbb{Q})H$, and they are trace preserving.
\end{lemma}

{\bf Proof.}
Since $(I-\mathbb{Q})N(I-\mathbb{Q})=N_{11}=0$, we have that 
$(I-\mathbb{Q})N=(I-\mathbb{Q})N\mathbb{Q}=N_{12}=K_{12}=(I-\mathbb{Q})K\mathbb{Q}$.
This leads to
$$
\begin{aligned}
(I-\mathbb{Q})L 
& = (I-\mathbb{Q})K - (I-\mathbb{Q})N\Phi 
= (I-\mathbb{Q})K - (I-\mathbb{Q})K\mathbb{Q}\Phi 
= (I-\mathbb{Q})K(I-\mathbb{Q}\Phi) 
\\
& = (I-\mathbb{Q})\Phi(I-\mathbb{Q}\Phi)^{-1} 
= (I-\mathbb{Q})H,
\end{aligned}
$$
where in the last equalities we have used \eqref{HK}. That they are trace preserving follows from Remark~\ref{finrec}.
\qed

The previous results lead directly to the MHTF for a state orthogonal to the arrival subspace.

\begin{theorem} \label{mhtf} 
(MHTF for positive maps and orthogonal states). Let $\Phi: M_n\to M_n$ denote an irreducible, positive, trace preserving map and $V$ a nontrivial subspace of the pure state Hilbert space $\HH_n$. If $Z$ is the associated fundamental map \eqref{FMPM}, $K$ is the mean hitting time map \eqref{HK} and $D$ is its diagonal part given in \eqref{DKL}, we have that for every $\psi\in V$ and $\phi\in V^\perp$,
\beq \label{mhtff1}
\tau(\phi\to V) = \Tr\Big((DZ)_{11}\rho_\psi-(DZ)_{12}\rho_\phi\Big)
= \Tr\Big(K_{11}(Z_{11}\rho_\psi-Z_{12}\rho_\phi)\Big).
\eeq
In particular, the quantity $\Tr((DZ)_{11}\rho_\psi)$ is independent of the choice of $\psi\in V$.
\end{theorem}

\begin{remark} 
In words, the meaning of the above result is: the mean time of first visit to a subspace is an information which can be obtained essentially in terms of the fundamental map $Z$ of the dynamics and the mean return time map $K_{11}$ for such subspace.
\end{remark}

\begin{remark} \label{mixed}
The monitored notions of hitting probability and mean hitting time make for an arbitrary initial state as much sense as for an initial pure state. The resulting expressions for these quantities follow by substituting $\rho_\phi$ by an arbitrary density matrix $\rho$ in \eqref{HKlim} and Remark~\ref{finrec}. It is straightforward to see that the MHTF is still true under this substitution if one reinterprets the condition $\phi\in V^\bot$ as $\mathbb{Q}\rho=\rho$. Similarly, the MHTF holds when substituting $\rho_\psi$ with $\psi\in V$ by an arbitrary density $\rho'$ such that $\mathbb{P}\rho'=\rho'$. These substitutions also do not alter the validity of the next results, Lemma~\ref{cfslemma} and Theorem~\ref{mhtfg}.
\end{remark}

{\bf Proof of Theorem \ref{mhtf}.} 
From \eqref{tauK}, by using the maps defined in \eqref{DKL}, we find that, for every $\psi\in V$ and $\phi\in V^\bot$,
$$
\tau(\phi\to V) = \Tr(N_{12}\rho_\phi)
= \Tr[(DZ)_{11}\rho_\psi-(DZ)_{12}\rho_\phi] + \Tr[(LZ)_{12}\rho_\phi-(LZ)_{11}\rho_\psi].
$$
On the other hand, since $Z$ and $(I-\mathbb{Q})L$ are trace preserving (see lemmas \ref{omega1lema} and \ref{Ltr}), 
$$
\Tr[(LZ)_{11}\rho_\psi] = \Tr[(I-\mathbb{Q})LZ\rho_\psi] = \Tr(\rho_\psi) = 1,
\quad
\Tr[(LZ)_{12}\rho_\phi] = \Tr[(I-\mathbb{Q})LZ\rho_\phi] = \Tr(\rho_\phi) = 1. 
$$
The theorem follows from the above two equalities.

\qed

The MHTF given by Theorem~\ref{mhtf} is restricted to states $\phi$ which are orthogonal to the arrival subspace $V$. It is possible to extend it to non-orthogonal states by using a general relation between the mean hitting time for a pure state $\phi$ and for the state obtained after the first step of the monitored evolution, i.e. the trace normalization of $\mathbb{Q} \Phi \rho_\phi$ (see Remark~\ref{mixed}).

\begin{lemma} (Conditioning on the first step for positive maps). \label{cfslemma} 
Let $\Phi$ be a positive trace preserving map on $M_n$, $V$ a subspace of the pure state Hilbert space $\HH_n$ and $\phi\in\HH_n$. Then,
\beq \label{CFS}
\tau(\phi\to V) = 
\begin{cases}
1, & \text{ if } \mathbb{Q}\Phi\rho_\phi=0,
\\
1 + \Tr(\mathbb{Q}\Phi\rho_\phi) \, \tau\left(\hat\rho_\phi \to V\right), \quad
\displaystyle \hat\rho_\phi = \frac{\mathbb{Q}\Phi\rho_\phi}{\Tr(\mathbb{Q}\Phi\rho_\phi)},
& \text{ if } \mathbb{Q}\Phi\rho_\phi\ne0.
\end{cases}
\eeq
\end{lemma}

{\bf Proof.} This is a generalization of a result appearing in \cite{oqwmhtf1}. From Remark~\ref{finrec} we already known that $\pi(\phi\to V)=1$. If $\mathbb{Q}\Phi\rho_\phi=0$, \eqref{probPOS} yields $\tau(\phi\to V) = \Tr(\mathbb{P}\Phi\rho_\phi) = \Tr(\Phi\rho_\phi) - \Tr((\mathbb{Q}+\mathbb{R})\Phi\rho_\phi) = \Tr(\rho_\phi) = 1$, since $\Phi$ is trace preserving and $\mathbb{R}\Phi\rho_\phi$ is traceless. Otherwise, we have
$$
\begin{aligned}
\tau(\phi\to V)-1 & = \tau(\phi\to V)-\pi(\phi\to V) = 
\sum_{n\geq 1} n \Tr(\mathbb{P} \Phi (\mathbb{Q} \Phi )^{n-1} \rho_\phi) 
- \sum_{n\geq 1} \Tr(\mathbb{P} \Phi (\mathbb{Q} \Phi )^{n-1} \rho_\phi)
\\
& = \sum_{n\geq 2} (n-1) \Tr(\mathbb{P} \Phi (\mathbb{Q} \Phi )^{n-1} \rho_\phi) 
= \sum_{n\geq 1} n\Tr(\mathbb{P} \Phi (\mathbb{Q} \Phi )^{n-1}\mathbb{Q} \Phi \rho_\phi) 
= \Tr(\mathbb{Q}\Phi\rho_\phi) \, \tau(\hat\rho_\phi \to V).
\end{aligned}
$$
\qed

Now we can combine the previous lemma with the version of Theorem~\ref{mhtf} for arbitrary density matrices pointed out in Remark~\ref{mixed}. The result is a MHTF for initial states $\phi$ which are not necessarily orthogonal to the arrival subspace $V$. 

\begin{theorem} \label{mhtfg} 
(MHTF for positive maps and non-orthogonal states). Under the conditions of Theorem~\ref{mhtf}, if $\phi\in\HH_n$ is arbitrary and $\hat\rho_\phi$ is given by \eqref{CFS}, then
$$
\tau(\phi\to V) = 
\begin{cases}
1, & \text{ if } \mathbb{Q}\Phi\rho_\phi=0,
\\
1 + \Tr(\mathbb{Q}\Phi\rho_\phi) \Tr\Big(K_{11}(Z_{11}\rho_\psi-Z_{12}\hat\rho_\phi)\Big),
& \text{ if } \mathbb{Q}\Phi\rho_\phi\ne0.
\end{cases}
$$
Equivalently,
$$
\tau(\phi\to V) = 
1 + \Tr(K_{11}Z_{11}\rho_\psi)\Tr(\mathbb{Q}\Phi\rho_\phi) 
- \Tr(K_{11}Z_{12}\mathbb{Q}\Phi\rho_\phi).
$$
\end{theorem}

The following examples will illustrate the versions of the MHTF for positive maps given in theorems \ref{mhtf} and \ref{mhtfg}.


\bex ($\dim V=1$) \label{ex1} 
Consider the positive map on $M_2$ given by
$$
\Phi(X)=LXL^*+RXR^*, \qquad 
L = \frac{1}{\sqrt{3}}\begin{bmatrix} 1 & 1 \\ 0 & 1 \end{bmatrix}, \qquad 
R = \frac{1}{\sqrt{3}}\begin{bmatrix} 1 & 0 \\ -1 & 1\end{bmatrix},
$$
which has the matrix representation
$$
\lceil\Phi\rceil = L\otimes\ov{L}+R\otimes \ov{R} = \frac{1}{3}
\begin{bmatrix} 
2 & 1 & 1 & 1 \\ -1 & 2 & 0 & 1 \\ -1 & 0 & 2 & 1 \\ 1 & -1 & -1 & 2
\end{bmatrix}.
$$
It is clear that this is a quantum channel, that is, a completely positive, trace preserving map since $L^*L+R^*R=I_2$. Moreover, it is unital, i.e., $\Phi(I_2)=I_2$, and irreducible, because the unique invariant state is $\pi=\frac{1}{2}I_2$, which is strictly positive \cite{carbone2}. The matrix representation of $\Omega=|\pi\rangle\langle I_2|$ is 
$$
\lceil\Omega\rceil = vec(\pi) \, vec(I_2)^* = \frac{1}{2}
\begin{bmatrix} 
1 & 0 & 0 & 1 \\ 0 & 0 & 0 & 0 \\ 0 & 0 & 0 & 0 \\ 1 & 0 & 0 & 1
\end{bmatrix},
\quad 
\lceil Z\rceil = (I_4-\lceil\Phi\rceil+\lceil\Omega\rceil)^{-1} 
= \frac{1}{4}
\begin{bmatrix} 
3 & 2 & 2 & 1 \\
-2 & 8 & -4 & 2 \\ 
-2 & -4 & 8 & 2 \\ 
1 & -2 & -2 & 3
\end{bmatrix}.
$$

\medskip

Let us choose initial and arrival states, respectively, as
$$
\phi = \frac{1}{\sqrt{2}} \begin{bmatrix} 1 \\ -1 \end{bmatrix},
\qquad
\psi = \frac{1}{\sqrt{2}} \begin{bmatrix} 1 \\ 1 \end{bmatrix}.
$$
The corresponding densities give also the orthogonal projections,
$$
\begin{aligned}
& P = \rho_\psi = \frac{1}{2}\begin{bmatrix} 1 & 1 \\ 1 & 1 \end{bmatrix},
& \qquad &
Q = I_2-P = \rho_\phi = \frac{1}{2}\begin{bmatrix} 1 & -1 \\-1 & 1\end{bmatrix},
\\ 
& \lceil\mathbb{P}\rceil = \rho_\psi \otimes \rho_\psi 
= \frac{1}{4} 
\begin{bmatrix} 
1 & 1 & 1 & 1 \\ 1 & 1 & 1 & 1 \\ 1 & 1 & 1 & 1 \\ 1 & 1 & 1 & 1 
\end{bmatrix},
& &  
\lceil\mathbb{Q}\rceil = \rho_\phi \otimes \rho_\phi 
= \frac{1}{4} 
\begin{bmatrix} 
1 & -1 & -1 & 1 \\ -1 & 1 & 1 & -1 \\ -1 & 1 & 1 & -1 \\ 1 & -1 & -1 & 1 
\end{bmatrix}.
\end{aligned}
$$
The mean hitting time map $K$ can be obtained via \eqref{HK}, 
$$
\lceil K\rceil = \lceil\Phi\rceil 
(I_4-\lceil\mathbb{Q}\rceil\lceil\Phi\rceil)^{-2} =
\frac{1}{6}
\begin{pmatrix}
39 & -12 & -12 & 9\\
-72 & 32 & 28 & -12\\
-72 & 28 & 32 & -12\\
177 & -72 & -72 & 39
\end{pmatrix},
$$
hence
$$
\lceil K_{12}\rceil = 
(I_4-\lceil\mathbb{Q}\rceil) \lceil K\rceil \lceil\mathbb{Q}\rceil 
= \frac{3}{2}
\begin{bmatrix}
-3 & 3 & 3 & -3 \\ 1 & -1 & -1 & 1 \\ 1 & -1 & -1 & 1 \\ 5 & -5 & -5 & 5
\end{bmatrix},
$$
which yields
$$
\lceil K_{12}\rceil \, vec(\rho_\phi) = \begin{bmatrix} -9 \\ 3 \\ 3 \\ 15 \end{bmatrix} 
\quad \Rightarrow \quad 
\tau(\phi\to\psi) = \Tr(K_{12}\phi) = -9 + 15 = 6.
$$
Following similar calculations, we have
$$
\Tr((DZ)_{11}\rho_\psi)=\Tr(K_{11}Z_{11}\rho_\psi)=4,
\qquad
\Tr((DZ)_{12}\rho_\phi)=\Tr(K_{11}Z_{12}\rho_\phi)=-2,
$$
from which we obtain the same mean hitting time by the MHTF \eqref{mhtff1}, as expected.

\medskip

Consider now the initial state
$$
\chi = \begin{bmatrix} 0 \\ 1 \end{bmatrix},
$$
which gives 
$$
\lceil \mathbb{P} \rceil \lceil K \rceil \, vec(\rho_\chi) 
= \begin{bmatrix} 1 \\ 1 \\ 1 \\ 1 \end{bmatrix} 
\quad \Rightarrow \quad
\tau(\chi\to\psi) = \Tr(\mathbb{P}K\rho_\chi) = 2.
$$
Since $\chi$ is not orthogonal to $\psi$, to compare with the calculation using the MHTF we should use the version given in Theorem~\ref{mhtfg}. Bearing in mind that 
$$
\lceil \mathbb{Q} \rceil \lceil \Phi \rceil \, vec(\rho_\chi) 
= \frac{1}{12} \begin{bmatrix} 1 \\ -1 \\ -1 \\ 1 \end{bmatrix}
\quad \Rightarrow \quad
\mathbb{Q} \Phi \rho_\chi = \frac{1}{6} \rho_\phi
\quad \Rightarrow \quad
\Tr(\mathbb{Q} \Phi \rho_\chi) = \frac{1}{6}, \quad 
\hat\rho_\chi = \rho_\phi,
$$
Lemma~\ref{cfslemma} and Theorem~\ref{mhtfg} permit us to write
$$
\tau(\chi\to\psi) = 1 + \frac{1}{6} \tau(\phi\to\psi) 
= 1 + \frac{1}{6} (\Tr(K_{11}Z_{11}\rho_\psi) - \Tr(K_{11}Z_{12}\rho_\phi)) 
= 1 + \frac{4}{6} - \Big(-\frac{2}{6}\Big) = 2,
$$
in agreement with the previous calculation.
\eex
\qee

\bex ($\dim V=2$) \label{ex2} 
For $0<a<1$ and $b=\sqrt{1-a^2}$ let 
$$
V_1 = \begin{bmatrix} 
a & 0 & 0 & b \\ 0 & 0 & 0 & 0 \\ 0 & 0 & 0 & 0 \\ 0 & 0 & 0 & 0
\end{bmatrix},
\quad 
V_2 = \begin{bmatrix} 
0 & 0 & 0 & 0 \\ -b & 0 & 0 & a \\ 0 & 0 & 0 & 0 \\ 0 & 0 & 0 & 0 
\end{bmatrix},
\quad
V_3 = \frac{1}{\sqrt{2}} \begin{bmatrix} 
0 & 0 & 0 & 0 \\ 0 & 0 & 0 & 0 \\ 0 & 1 & 1 & 0 \\ 0 & 0 & 0 & 0 
\end{bmatrix},
\quad
V_4 = \frac{1}{\sqrt{2}} \begin{bmatrix} 
0 & 0 & 0 & 0 \\ 0 & 0 & 0 & 0 \\ 0 & 0 & 0 & 0 \\ 0 & 1 & -1 & 0 
\end{bmatrix}.
$$
The completely positive map on $M_4$ given by $\Phi(X)=\sum_{i=1}^4 V_i XV_i^*$ is trace-preserving and irreducible, the unique invariant state being $\pi=\frac{1}{4}I_4$. One may calculate $\Omega=|\pi\rangle\langle I_4|$ from which the map $Z$ is immediately obtained (for brevity, we omit its explicit expression).

\medskip 

Let 
$$
\psi_1 = \begin{bmatrix} 0 \\ 0 \\ 1 \\ 0 \end{bmatrix},
\qquad 
\psi_2 = \begin{bmatrix} 0 \\ 0 \\ 0 \\ 1 \end{bmatrix},
\qquad
\phi=\begin{bmatrix} 1 \\ 0 \\ 0 \\ 0 \end{bmatrix}.
$$
We will consider $V=\spn\{\psi_1, \psi_2\}$ as a two-dimensional arrival subspace and $\phi$ as the initial state. Hence, the corresponding projections are $P=\rho_{\psi_1}+\rho_{\psi_2}$ and $Q=I_4-P$. Once again, from $\lceil\mathbb{P}\rceil =P\otimes P$ and $\lceil\mathbb{Q}\rceil =Q\otimes Q$, routine calculations lead us to the order 16 matrix
$$
\lceil K\rceil = \lceil\Phi\rceil (I_{16}-\lceil\mathbb{Q}\rceil\lceil\Phi\rceil)^{-2} 
= \begin{bmatrix} A_1 & A_2 \\ A_3 & A_4 \end{bmatrix},
$$
where
$$
\begin{aligned}
& A_1 = \begin{bmatrix}
\frac{a^2}{b^4} & 0 & 0 & \frac{a}{b^3} & 0 & 0 & 0 & 0 \\[1pt]
0 & 0 & 0 & 0 & 0 & 0 & 0 & 0 \\
0 & 0 & 0 & 0 & 0 & 0 & 0 & 0 \\
0 & 0 & 0 & 0 & 0 & 0 & 0 & 0 \\
0 & 0 & 0 & 0 & 0 & 0 & 0 & 0 \\
\frac{1}{b^2} & 0 & 0 & \frac{a}{b} & 0 & 0 & 0 & 0 \\[1pt]
0 & 0 & 0 & 0 & 0 & 0 & 0 & 0 \\
0 & 0 & 0 & 0 & 0 & 0 & 0 & 0 
\end{bmatrix},
& \quad & 
A_2 = \begin{bmatrix}
0 & 0 & 0 & 0 & \frac{a}{b^3} & 0 & 0 & \frac{1}{b^2} \\[1pt]
0 & 0 & 0 & 0 & 0 & 0 & 0 & 0 \\
0 & 0 & 0 & 0 & 0 & 0 & 0 & 0 \\
0 & 0 & 0 & 0 & 0 & 0 & 0 & 0 \\
0 & 0 & 0 & 0 & 0 & 0 & 0 & 0 \\
0 & 0 & 0 & 0 & \frac{a}{b} & 0 & 0 & 2 \\[1pt]
0 & 0 & 0 & 0 & 0 & 0 & 0 & 0 \\
0 & 0 & 0 & 0 & 0 & 0 & 0 & 0 
\end{bmatrix},
\\
& A_3 = \begin{bmatrix}
0 & 0 & 0 &0 & 0 & 0 & 0 & 0 \\
0 & 0 & 0 & 0 & 0 & 0 & 0 & 0 \\
\frac{1+b^2}{2b^2} & 0 & 0 & \frac{a}{2b} & 0 & \frac{1}{2} & \frac{1}{2} & 0 \\[1pt]
0 & 0 & 0 & 0 & 0 & 0 & 0 & 0 \\
0 & 0 & 0 & 0 & 0 & 0 & 0 & 0 \\
0 & 0 & 0 & 0 & 0 & 0 & 0 & 0 \\
0 & 0 & 0 & 0 & 0 & 0 & 0 & 0 \\
\frac{1+b^2}{2b^2} & 0 & 0 & \frac{a}{2b} & 0 & \frac{1}{2} & -\frac{1}{2} & 0 
\end{bmatrix},
& & 
A_4 = \begin{bmatrix}
0 & 0 & 0 & 0 & 0 & 0 & 0 & 0 \\
0 & 0 & 0 & 0 & 0 & 0 & 0 & 0 \\
0 & \frac{1}{2} & \frac{1}{2} & 0 & \frac{a}{2b} & 0 & 0 & \frac{3}{2} \\[1pt]
0 & 0 & 0 & 0 & 0 & 0 & 0 & 0 \\
0 & 0 & 0 & 0 & 0 & 0 & 0 & 0 \\
0 & 0 & 0 & 0 &0 & 0 & 0 & 0 \\
0 & 0 & 0 & 0 & 0 & 0 & 0 & 0 \\
0 & -\frac{1}{2} & \frac{1}{2} & 0 & \frac{a}{2b} & 0 & 0 & \frac{3}{2}
\end{bmatrix}.
\end{aligned}
$$
This yields
$$
K_{12} \rho_\phi = 
\begin{bmatrix} 
0 & 0 & 0 & 0\\ 0 & 0 & 0 & 0 \\ 0 & 0 & \frac{1+b^2}{2b^2} & 0 \\ 0 & 0 & 0 & \frac{1+b^2}{2b^2}
\end{bmatrix} 
\quad \Rightarrow \quad 
\tau(\phi\to V) = \Tr(K_{12}\rho_\phi) = \frac{1+b^2}{b^2} = 1 + \frac{1}{b^2}.
$$

To compare wiht the result via MHTF, consider an arbitrary density $\rho$ such that $\mathbb{P}\rho=\rho$, i.e., 
\begin{equation} \label{gdex}
\rho = \begin{bmatrix}
0 & 0 & 0 & 0 \\
0 & 0 & 0 & 0 \\
0 & 0 & x & y \\
0 & 0 & \overline{y} & 1-x \\
\end{bmatrix},
\qquad
0 \le x \le 1, \qquad |y| \le \sqrt{x(1-x)}.
\end{equation}
Then, proceeding as previously by using matrix representations yields  
$$
\Tr((DZ)_{11}\rho) = \Tr(K_{11}Z_{11}\rho) 
= \frac{1+6b^2}{4b^2},
\qquad
\Tr((DZ)_{12}\rho_\phi) = \Tr(K_{11}Z_{12}\rho_\phi) 
= \frac{2b^2-3}{4b^2}.
$$
Using the MHTF \eqref{mhtff1} and taking into account Remark~\ref{mixed}, this gives the same mean hitting time $\tau(\phi\to V)$ already obtained. Note also that the trace $\Tr((DZ)_{11}\rho)$ is independent of the chosen density \eqref{gdex}, illustrating a general property pointed out in Theorem~\ref{mhtf} and extended in Remark~\ref{mixed}.  

\medskip

Now consider the initial state $\chi$ given by
$$
\chi = \frac{1}{\sqrt{2}}\begin{bmatrix} 1 \\ 0\\0\\1 \end{bmatrix}.
$$
In this case one obtains  
$$
\mathbb{P} K \rho_\chi 
= \begin{bmatrix} 
0 & 0 & 0 & 0 \\ 0 & 0 & 0 & 0 \\ 0 & 0 & c & 0 \\ 0 & 0 & 0 & c
\end{bmatrix},
\quad
c = \frac{1+2ab+4b^2}{4b^2} = 1 + \frac{a}{2b} + \frac{1}{4b^2}
\quad \Rightarrow \quad
\tau(\chi\to V) = \Tr(\mathbb{P}K\rho_\chi) = 2c.
$$
As $\chi$ is not orthogonal to the arrival subspace $V$, we should compare this result with that one obtained using Theorem~\ref{mhtfg}. Since
$$
\mathbb{Q} \Phi \rho_\chi 
= \begin{bmatrix} 
\frac{1}{2}+ab & 0 & 0 & 0 \\ 0 & \frac{1}{2}-ab & 0 & 0 \\ 0 & 0 & 0 & 0 \\ 0 & 0 & 0 & 0
\end{bmatrix}
\quad \Rightarrow \quad
\Tr(\mathbb{Q} \Phi \rho_\chi) = 1, \quad 
\hat\rho_\chi = \mathbb{Q} \Phi \rho_\chi,
$$
using the density \eqref{gdex}, Lemma~\ref{cfslemma} and Theorem~\ref{mhtfg} imply
$$
\tau(\chi\to V) = 1 + \tau(\mathbb{Q} \Phi \rho_\chi\to V) 
= 1 +  \Tr(K_{11}Z_{11}\rho) - \Tr(K_{11}Z_{12}\mathbb{Q}\Phi\rho_\chi) 
= 1 + \left(\frac{1+6b^2}{4b^2}\right) + \left(\frac{1+4ab-2b^2}{4b^2}\right),
$$
which equals $2c$, in agreement with the previous calculation.
\eex
\qee

\section{The classical MHTF revisited}\label{sec5}

To understand that Theorem~\ref{mhtf} is a right generalization of the classical MHTF, we must show that \eqref{mhtff1} becomes \eqref{mhtf_classica} in the classical setting. Nevertheless, this is not the only purpose of this section. The wide generality of the MHTF obtained in this work will allow us to obtain classical MHTF which generalize \eqref{mhtf_classica} in two senses: either by assuming only the knowledge of the probabilistic distribution for the initial state, or by enlarging the arrival state to a subset of states.

\medskip

The first step is to describe classical Markov chains in terms of positive trace preserving maps. Let $\Pr=(p_{ij})$ be the stochastic matrix of an irreducible finite Markov chain with $n$ states indexed by $i=1,2,\dots,n$. We can associate with this chain a Hilbert space $\HH=\spn\{|i\rangle : i=1,2,\dots,n\}$ spanned by an orthonormal set of vectors indexed by the states of the chain. Then, using the ket-bra operators $|i\rangle\langle j|\phi = \langle j|\phi\rangle |i\rangle$ on $\HH$, we can define a completely positive map $\Phi$ on $M_n$,
\begin{equation} \label{Phic}
\Phi = \sum_{i,j} p_{ij} |i\>\<j| \cdot |j\>\<i|,
\end{equation}
which acts as follows on every matrix $(x_{ij})\in M_n$, rewritten as $X = \sum_{i,j} x_{ij}|i\rangle\langle j|$,
\begin{equation} \label{Phiclas}
\Phi X = \sum_{i,j} p_{ij} x_{jj} \rho_i, \qquad \rho_i=|i\>\<i|.
\end{equation}
This means that $\Phi$ acts non-trivially only on the subspace $D_n \subset M_n$ of diagonal matrices $X=\sum_i x_{ii}\rho_i$, i.e. $\Phi(M_n)\subset D_n$. Hence, $D_n$ is an invariant subspace for $\Phi$, and the action of $\Phi$ on $D_n$ mimics that of the Markov chain: $x \mapsto \Pr x$ with $x=(x_{ii})$ the column vector given by the diagonal of $X$. 

\medskip
  
The map $\Phi$ is trace preserving because $\Pr$ is stochastic. The irreducibility of $\Pr$ implies the existence of a unique invariant distribution $\pi=(\pi_i)$ for $\Pr$ with $\pi_i>0$ for all $i$. Thus, the strictly positive density $\sum_i\pi_i\rho_i$ is the unique invariant state for $\Phi$, which implies the irreducibility of the completely positive map $\Phi$  \cite{carbone2}. We will assume that the positive map $\Phi$ defined by \eqref{Phic} is restricted to $D_n$, so that it is equivalent to the Markov chain given by $\Pr$.

\medskip

Consider an initial state $i$ and an arrival state $j$. If $j\ne i$, they are represented by orthogonal states in the auxiliary Hilbert space $\HH$, hence Theorem~\ref{mhtf} applies directly to the mean hitting time $\tau(i \to j)$, the arrival subspace being $V=\spn\{|j\rangle\}$. The corresponding orthogonal projections on $\HH$ are $P=\rho_j$ and $Q=\sum_{i\ne j}\rho_i$, while the orthogonal projections $\mbb{P}=P \cdot P$ and $\mbb{Q}=Q \cdot Q$ on $D_n$ satisfy $\mbb{P}+\mbb{Q}=I$ because the traceless projection $\mbb{R} = I-\mbb{P}-\mbb{Q} = P \cdot Q + Q \cdot P$ vanishes on the subspace $D_n$ which supports the dynamics. 

\medskip

The fundamental map $Z=\sum_{k,l}Z_{kl}|k\rangle\langle l|\cdot|l\rangle\langle k|$ acts on $D_n$ similarly to \eqref{Phiclas}, i.e. 
$$
Z X = \sum_{k,l} Z_{kl} x_{ll} \rho_k 
\quad \Rightarrow \quad Z \rho_l = \sum_k Z_{kl} \rho_k
\quad \Rightarrow \quad \mbb{P} Z \rho_l = Z_{jl} \rho_j.
$$
In consequence, the terms involved in the MHTF \eqref{mhtff1} for $\tau(i\to j)$ read as
\begin{equation} \label{trtr}
\Tr(\mbb{P}K\mbb{P}Z\rho_j) = Z_{jj} \Tr(\mbb{P}K\rho_j) = Z_{jj} \, \tau(j\to j),
\qquad 
\Tr(\mbb{P}K\mbb{P}Z\rho_i) = Z_{ji} \Tr(\mbb{P}K\rho_j) = Z_{ji} \, \tau(j\to j).
\end{equation}
Using these equalitites, the MHTF \eqref{mhtff1} becomes the classical MHTF \eqref{mhtf_classica} bearing in mind Kac's lemma \cite{durrett,kac}, which states that $\tau(j\to j)=1/\pi_j$. 

\medskip

Suppose now that we only know the probability $x_i$ that the initial state is $i$. This defines an initial formal state given by a probability distribution $x=(x_i)$ satisfying $x_i\ge0$ and $\sum_i x_i=1$. In terms of the ingredients for the positive map description of the chain, this initial state is identified with the density matrix $\rho = \sum_i x_i \rho_i$. In this situation, considering an arrival state $j$ means that, as previously, $P=\rho_j$, $Q=\sum_{i\ne j}\rho_i$, $\mathbb{P}+\mathbb{Q}=I$ and $\mathbb{P}Z\rho_l=Z_{jl}\rho_j$. Then, Theorem~\ref{mhtfg} and Remark~\ref{mixed} yield the following expression for the mean time to reach the state $j$ subject to the initial distribution $x$, 
$$
\tau(x\to j) 
= 1 + \Tr(\mathbb{P}K\mathbb{P}Z\rho_j)\Tr(\mathbb{Q}\Phi\rho) 
- \Tr(\mathbb{P}K\mathbb{P}Z\mathbb{Q}\Phi\rho).
$$
Here $\Tr(\mathbb{P}K\mathbb{P}Z\rho_j)$ is given by \eqref{trtr}, while $\Phi\rho = \sum_i (\Pr x)_i\rho_i$ leads to
$$
\begin{gathered}
\Tr(\mathbb{Q}\Phi\rho) = \Tr(\Phi\rho)-\Tr(\mathbb{P}\Phi\rho) = 1-(\Pr x)_j,
\\
\Tr(\mathbb{P}K\mathbb{P}Z\mathbb{Q}\Phi\rho) 
= \sum_{i\ne j} (\Pr x)_i \Tr(\mathbb{P}K\mathbb{P}Z\rho_i) 
= \sum_{i\ne j} Z_{ji} (\Pr x)_i \Tr(\mathbb{P}K\rho_j)
= \sum_{i\ne j} Z_{ji} (\Pr x)_i \, \tau(j\to j). 
\end{gathered}
$$
Combining the above results we find that
$$
\begin{aligned}
\tau(x\to j) 
& = 1 + Z_{jj} \left(1-(\Pr x)_j\right) \tau(j\to j) 
- \sum_{i\ne j} Z_{ji} (\Pr x)_i \, \tau(j\to j)
\\
& = 1 + Z_{jj} \tau(j\to j) - \sum_i Z_{ji} (\Pr x)_i \, \tau(j\to j). 
\end{aligned}
$$
Therefore, using again Kac's lemma, we conclude the following result:

\begin{theorem} 
Let $\Pr=(p_{ij})$ be the stochastic matrix of an irreducible finite Markov chain with $n$ states indexed by $i=1,2,\dots,n$, invariant distribution $\pi=(\pi_i)$ and associated fundamental map $Z$. Then, given an initial probability distribution $x=(x_i)$, the mean time to reach the state $j$ is given by 
$$
\tau(x\to j) = 1 + \frac{Z_{jj} - (Z \Pr x)_j}{\pi_j}.
$$ 
\end{theorem}
This may be considered as the classical MHTF for an arbitrary initial probability distribution $x$.

\medskip

Consider now a subset of states $S \subset \{1,2,\dots,n\}$. A MHTF for the arrival subset $S$ should express the mean time $\tau(i\to S)$ of the first visit to $S$ starting from a state $i\notin S$ in terms of the mean return times $\tau(k\to S)$, $k\in S$, and the fundamental matrix $(Z_{kl})$. Such a MHTF should follow by applying Theorem~\ref{mhtf} with $V=\spn\{|k\rangle : k\in S\}$ and $P=\sum_{k\in S}\rho_k$ because $|i\rangle$ is orthogonal to $V$. The direct application of this theorem gives 
$$
\tau(i\to S) = \Tr(\mathbb{P}K\mathbb{P}Z\rho_j) - \Tr(\mathbb{P}K\mathbb{P}Z\rho_i),
\qquad j\in S, \quad i\notin S.
$$
Since now $\mathbb{P}Z\rho_l=\sum_{k\in S}Z_{kl}\rho_k$, we obtain
$$
\tau(i\to S) 
= \sum_{k\in S} (Z_{kj}-Z_{ki}) \Tr(\mathbb{P}K\rho_k) 
= \sum_{k\in S} (Z_{kj}-Z_{ki}) \, \tau(k\to S),
\qquad j\in S, \quad i\notin S,
$$
which is the classical MHTF for an arrival subset of states $S$, summarized below.
\begin{theorem}
For an irreducible finite Markov chain with $n$ states indexed by $i=1,2,\dots,n$, invariant distribution $\pi=(\pi_i)$ and fundamental map $Z$, the mean time to reach a subset of states $S \subset \{1,2,\dots,n\}$ starting at an initial state $i\notin S$ is given by
$$
\tau(i\to S)=\sum_{k\in S} (Z_{kj}-Z_{ki}) \, \tau(k\to S),
\qquad j\in S, \quad i\notin S.
$$
In particular, the sum $\sum_{k\in S} Z_{kj} \, \tau(k\to S)$ is independent of the chosen state $j\in S$.
\end{theorem}

\bigskip

{\bf Acknowledgments.} The first author acknowledges financial support from a CAPES/PROAP grant (Programa de Apoio \`a
P\'os-Gradua\c c\~ao - 2019) to PPGMat/UFRGS. The work of the second author is part of the I+D+i project MTM2017-89941-P funded by MCIN/ AEI/10.13039/501100011033/ and ERDF ``Una manera de hacer Europa'', the project UAL18-FQM-B025-A (UAL/CECEU/FEDER) and the project E48\_20R from Diputaci\'on General de Arag\'on (Spain) and ERDF ``Construyendo Europa desde Arag\'on''.

\end{document}